\documentclass[12pt]{iopart}
\usepackage{epsf}

\begin{document}

\title[Classical Coulomb three-body problem]
{Classical Coulomb three-body problem\\
in collinear eZe configuration}

\author{Mitsusada~M.~Sano}

\address{
Graduate School of Human and Environmental Studies\\
Kyoto University,\\
Sakyo, Kyoto, 606-8501, Japan
}
\date{7 Feb 2003}
\begin{abstract}
Classical dynamics of two-electron atom and ions 
H$^{-}$, He, Li$^{+}$, Be$^{2+}$, ... 
in collinear eZe configuration is investigated. 
We consider the case 
that the masses of all particles are finite. 
It is revealed that the mass ratio $\xi$ 
between nucleus and electron 
plays an important role for dynamical behaviour of these systems. 
With the aid of analytical tool and numerical computation, 
it is shown that thanks to large mass ratio $\xi$, 
classical dynamics of these systems is fully chaotic, 
probably hyperbolic. 
Experimental manifestation of this finding is also proposed. 
\end{abstract}
\submitto{\JPA}
\maketitle
\section{Introduction} \label{sec1}
Three-body problem in celestial mechanics is 
one of pioneer studies on chaotic phenomena and, 
at the same time, one of the oldest and the most famous problems in physics. 
First systematic work has been given by Poincar\'{e},  
for which he won the prize established by King Oscar II 
of Sweden and Norway(Poincar\'{e}, 1899). 
Main result of his work is that 
the motion of three particles is very complicated
(i.e., what we now call ``chaos'') 
and then it is generally hard to solve it practically. 
After Poincar\'{e}, many mathematicians and physicists 
contributed to this problem
(See, for instance, the book(Siegel and Moser, 1971) 
as mathematical literature.). 
In addition, 
innovation of computers assisted physicists to investigate this problem. 
However, our understanding of three-body problem is far from complete. 
The investigation is still going on. 

Three-body problem also exists in Coulomb systems. 
The most popular Coulomb three-body system is helium atom 
which consists of one nucleus and two electrons. 
In early stage of making quantum mechanics, 
many physicists did not know what to do with the spectrum of helium atom, 
in spite of success of explaining of the spectrum of hydrogen atom. 
For hydrogen atom, stable periodic orbits correspond to 
eigen energies of it. 
This correspondence is now called 
the Bohr-Sommerfeld(BS) quantization. 
However, they could not explain the spectrum of helium atom 
by the BS-quantization scheme. 
Einstein extended the BS-quantization 
to higher-dimensional classically integrable case
(now we call it the Einstein-Brioullin-Keller(EBK) quantization.) 
and pointed out indirectly this problem of helium atom
(i.e., not integrable case) referring to the work by Poincar\'{e}
(Einstein, 1917). 
About fifty years later, Gutzwiller succeeded to derive 
an approximate quantization condition for hyperbolic cases. 
His formula expresses the density of states 
in terms of unstable periodic orbits, 
which is now called the Gutzwiller trace formula
(Gutzwiller, 1971, 1990). 
In early 90's, several physicists applied the Gutzwiller trace formula 
to hydrogen negative ion(Gaspard and Rice, 1993) and 
helium atom(Ezra et al., 1991) in collinear configuration, 
i.e., restricted Coulomb three-body problem. 
Series of these studies showed that 
these atom and ions in collinear eZe configuration 
are fully chaotic, probably hyperbolic and 
the Gutzwiller trace formula yields nice approximate values 
for series of eigen energies with angular momentum ${\bf L}=0$ 
numerically(See also the review(Tanner, Richter and Rost, 2000).). 
Here ``e'' and ``Z'' represent electron and nucleus, respectively. 
Thus ``eZe'' stands for the order of particles on a line. 
Their work has left a question 
why hydrogen negative ion and helium atom 
in collinear eZe configuration are fully chaotic. 
We used the term ``fully chaotic'' in the meaning 
that almost all periodic orbits are hyperbolic. 
Thus, we used the term ``fully chaotic'' 
for the system which is almost chaotic, but posseses small tiny island. 
In addition, we shall use the term ``hyperbolic'' in the meaning that 
all periodic orbits are hyperbolic. 

It was numerically shown that if the mass of nucleus is infinite for eZe 
collinear configuration, 
the symbolic description of it is complete for the negative energy
$E<0$ and $Z\geq 1$(Bai, Gu and Yuan, 1998). This result suggested that 
for the inifinite nucleus mass, the dynamics is hyperbolic. 

In this paper, 
we consider the case that all masses are finite 
and numerically confirm that the dyanmics of two-electron atom and ions 
in collinear eZe configuration, 
i.e., H$^{-}$, He, Li$^{+}$, Be$^{2+}$, ... is hyperbolic. 
Parameters of these systems are the mass ratio $\xi=m_{n}/m_{e}$ 
and the charge $Z$ of nucleus, where $m_{n}$($m_{e}$) are 
the mass of nucleus(electron), respectively. 
We employ numerical computation for these systems 
with the aid of analytical tool(the triple collision manifold(TCM))
(McGehee, 1974) and symbolic dynamics(Tanikawa and Mikkola, 2000) 
from celestial mechanics. 
The TCM is a manifold which is a set of the initial or final conditions 
of the triple collision orbit, 
i.e., thus just the triple collision points. 
The flow on the TCM does not have physical reality, 
because the TCM corresponds just triple collisions. 
However, the flow on the TCM affects the flow near the TCM 
because of the continuous property of the solution 
for the equations of motion. 
Thus, we can study the behaviour near the triple collision 
from the flow on the TCM. 
It is shown that the mass ratio $\xi$ plays an important role 
for dynamical behaviour of these systems. 
In precise, sufficiently large mass ratio $\xi$ makes 
the system be fully chaotic.  
Therefore it is revealed that all of these systems are fully chaotic, 
probably hyperbolic, since the mass ratio is sufficiently large, 
e.g., $m_{p}/m_{e} \approx 1840$, where $m_{p}$ is the mass of proton. 

We confirm this result as follows: 
First, we elucidate the $\xi$-dependence of the local structure 
near the triple collision by examining the flow on the TCM. 
In particular, we investigate the stable(or unstable) manifold of 
two critical points on the TCM. 
By this analysis, we show the change of the triple collision orbit 
when $\xi$ is increased. 
Let us denote the winding number of the stable(or unstable) manifold 
(on the TCM) of the critical point around the body of the TCM by ${\cal N}$. 
Later we shall define the body of the TCM precisely. 
It is shown that when $\xi \rightarrow 0$ , 
${\cal N}$ becomes infinite 
and, in the opposite limit, when $\xi$ is increased, 
${\cal N}$ is decreased and saturated to certain values. 
Second, we examine the morphology of the Poincar\'{e} surface of section 
using the associated symbolic dynamics. 
The symbolic dynamics is naturally introduced by labeling 
the double collision between the particles $1$ and $2$($2$ and $3$) 
by symbol $1$($2$), respectively. 
The position of the triple collision orbits 
on the Poincar\'{e} surface of section is specified. 
When $(Z,\xi)=(1,1)$, 
there is a torus on the Poincar\'{e} surface of section, 
whose corresponding sequence of symbols is $\dots 12121212 \dots$. 
The triple collision orbits forms curves 
on the Poincar\'{e} surface of section. 
We call these curves the {\em triple collision curves} ${\cal C}_{TC}$. 
In this case, ${\cal C}_{TC}$ crosses its reversed pair ${\cal C}^{t}_{TC}$ 
transversely except near the torus. 
${\cal C}^{t}_{TC}$ is mapped to ${\cal C}_{TC}$ by the symmetry operation 
due to the reverse operation. 
For helium $(Z,\xi)=(1,4m_{p}/m_{e})$, 
there is no torus. 
${\cal C}_{TC}$ crosses its pair ${\cal C}^{t}_{TC}$ 
transversely everywhere. 
From numerical calculation, 
${\cal C}_{TC}$(${\cal C}_{TC}^{t}$) is parallel to 
the stable manifolds(the unstable manifolds) in the Poincar\'{e} 
section for the Poincar\'{e} map, respectively. 
Therefore, 
this fact is an strong evidence 
that the stable and unstable manifolds transversely 
cross each other for the case of helium in collinear eZe configuration. 
In addition, we show the relationship between the structure of 
the stable and unstable manifolds of two critical points on the TCM 
and ${\cal C}_{TC}$ and ${\cal C}_{TC}^{t}$ on the Poincar\'{e} section. 
At this time, 
it is turned out 
that the change of the triple collision orbit on the TCM 
w.r.t $\xi$ is related to the collapse of the torus whose corresponding 
symbol sequence is $\dots 12121212 \dots$. 
The above facts strongly suggest that 
two-electron atom and ions in collinear eZe configuration is fully chaotic, 
probably hyperbolic. 

The organization of this paper is as follows. 
In \S \ref{sec2}, 
the Hamiltonian of the systems considered in this paper is introduced. 
Due to the attractive interaction between neighbouring particles, 
double collisions surely occur. 
Therefore, an regularization is required before numerical computation. 
We choose an algorithmic regularization not analytical regularization 
such as Kustaanheimo-Stiefel transformation or Levi-Civita transformation. 
In \S \ref{sec3}, the TCM is introduced. 
The $\xi$-dependence of the flow on the TCM is numerically investigated. 
In particular, the change of the triple collision orbit 
w.r.t. $\xi$ is shown. 
In \S \ref{sec4}, method of symbolic dynamics is applied to 
the systems for the case $Z=1$ and $\xi=1$ and the case of helium. 
In the former case, there is a torus whose symbol sequence is 
$\dots 12121212 \dots$. 
We examine the topological character of this system in detail. 
On the other hand, in the case of helium, 
there is no torus whose corresponding symbol sequence is 
$\dots 12121212 \dots$. In fact, the corresponding orbit for 
the symbol sequence $\dots 12121212 \dots$ 
is an isolated unstable periodic orbit. 
With the result of \S \ref{sec3}, 
what this observation means is considered. 
In \S \ref{sec5}, the results of this paper is summarized. 
\section{Hamiltonian and regularization} \label{sec2}
We consider three particles $1$, $2$ and $3$  
whose masses are $m_{1}=m_{e}$, $m_{2}=m_{n}=\xi m_{e}$, and $m_{3}=m_{e}$ 
and whose charges are $-e$, $Ze$, and $-e$, respectively. 
The mass ratio $\xi$ is given by $\xi = m_{n}/m_{e}$. 
The Hamiltonian for this system is 
\begin{equation}
H = \sum_{i=1}^{3} \frac{p_{i}^{2}}{2m_{i}} -  
\frac{Ze^{2}}{|q_{1}-q_{2}|} - \frac{Ze^{2}}{|q_{2}-q_{3}|} 
+ \frac{e^{2}}{|q_{1}-q_{3}|}.
\end{equation}
Now we employ the famous scaling for Coulomb systems. 
\begin{equation}
q_{i} = \alpha q_{i}', \;
p_{i} = \beta p_{i}', \; 
E = \gamma E' 
\end{equation}
with
\begin{equation}
\alpha =  \frac{Z}{m_{e}e^{2}|E|}, \;
\beta  =  |E|^{1/2}m_{e}e^{2},\; 
\gamma =  m_{e}e^{4}.
\end{equation}
$E$ is the value of the Hamiltonian $H$. 
After this scaling, we get the following Hamiltonian.
\begin{equation}
{\cal H} = 
\frac{p_{1}'^{2}}{2} + \frac{p_{2}'^{2}}{2\xi} + \frac{p_{3}'^{2}}{2}
- \frac{1}{|q_{1}'-q_{2}'|} - \frac{1}{|q_{2}'-q_{3}'|} 
+ \frac{1}{Z|q_{1}'-q_{3}'|} = E' = -1. 
\label{eq:H0}
\end{equation}
For simplicity, we set $E'=-1$. 
Now the particles are arranged in the order $q_{1}'\leq q_{2}'\leq q_{3}'$.
If we set $(Z,\xi)=(-1,1)$, the system is equivalent to the system 
which was investigated in gravitational three-body problem 
(Tanikawa and Mikkola, 2000). 
We set the total momentum to be zero and change the variables $q_{i}'$'s to 
\begin{equation}
Q_{1} = q_{2}' - q_{1}', \; 
Q_{2} = q_{2}', \; 
Q_{3} = q_{3}' - q_{2}'.
\end{equation}
with new momenta 
\begin{equation}
P_{1} = -p_{1}', \;
P_{2} =  p_{1}' + p_{2}' - p_{3}', \;
P_{3} =  p_{3}'.
\end{equation}
This canonical transformation is generated 
by the following generating function $W$.
\begin{equation}
W = P_{1} (q_{2}'-q_{1}') + P_{2} q_{2}' + P_{3}(q_{3}'- q_{2}'). 
\end{equation}
The final form of the Hamiltonian ${\cal H}$ is 
\begin{equation}
{\cal H} = 
\frac{P_{1}^{2}}{2\mu} + \frac{P_{2}^{2}}{2\mu} - \frac{P_{1}P_{2}}{\xi} 
- \frac{1}{Q_{1}} - \frac{1}{Q_{2}} + \frac{1}{Z(Q_{1}+Q_{2})} = -1,
\label{eq:H}
\end{equation}
with $\mu=\xi/(\xi + 1)$. 
After the canonical transformation, 
we replaced $(P_{3},Q_{3})$ by $(P_{2},Q_{2})$ for convenience. 
The parameters of this Hamiltonian is the charge $Z$ 
and the mass ratio $\xi$. 

The potential in Eq.(\ref{eq:H}) has singularities 
associated to double collisions of particles. 
Thus appropriate regularization is needed. 
To regularize them, in celestial mechanics a transformation such as 
Kustaanheimo-Stiefel transformation or Levi-Civita transformation 
is usually employed. 
However, an algorithmic regularization(Mikkola and Tanikawa, 1999) 
is used here. 
A merit of this choice is that 
for accurate numerical integration of equations of motion,  
the symplectic integrator method can be used(Yoshida, 1990). 
For usual analytical regularization,  
the symplectic integrator method can not be applied, 
since there are coupling terms of position and momentum in the Hamiltonian, 
i.e., the Hamiltonian after the transformation mentioned above 
is not a summation of the form ${\cal H}=T({\bf P})+V({\bf Q})$. 
We will use the sixth order symplectic integrator method in \S \ref{sec4}. 
After regularization, 
double collision is extended to just elastic collision and 
triple collision is not regularized in general. 
The interaction between outer particles and middle one is attractive, 
while the interaction between outer particles is repulsive. 
Thus neighbouring particles are always attracted and collide each other. 
Hence the trajectory of three particles is, in general, 
a sequence of collisions. 
Among collisions, triple collisions exist as very rare events. 
In fact, the measure of set of triple collisions would be zero. 
However, the triple collisions form 
the bone structure of dynamics of our systems 
which we will see in \S \ref{sec3} and \ref{sec4}. 

Here the symbolic coding of the orbit 
which will be extensively used in \S \ref{sec4} is introduced briefly. 
Let us denote the whole set of orbits in our systems by ${\cal O}$. 
The set ${\cal O}$ is an union of ${\cal O}_{DC}$ and ${\cal O}_{TC}$,
where ${\cal O}_{DC}$ is the set of orbits which only consist of double 
collisions and ${\cal O}_{TC}$ is the set of orbits 
which include triple collisions. 
\begin{equation}
{\cal O} = {\cal O}_{DC} \cup {\cal O}_{TC}.
\end{equation}
For an orbit of ${\cal O}_{DC}$, 
it is natural that 
the double collision with the particle $1$ and $2$($2$ and $3$) is 
labeled by symbol $1$($2$), respectively. 
Therefore, the symbol set for ${\cal O}_{DC}$ is ${\bf A}_{0}=\{ 1, 2 \}$.
Let us express an given orbit of ${\cal O}_{DC}$ 
as a sequence of symbols as follows:
\begin{equation}
{\bf n} = \dots n_{-2}n_{-1}\cdot n_{0} n_{1} n_{2} \dots,
\end{equation}
where $n_{i} \in {\bf A}_{0}, i\in {\bf Z}$. 
In order to describe an orbit of ${\cal O}_{TC}$ as a symbol sequence, 
we need another symbol, i.e., $0$. 
We set ${\bf A}=\{ 0,1,2 \}$. 
The triple collision orbit is started and/or ended by the triple collision. 
Let us label the triple collision by symbol $0$. 
The orbit which experience the triple collision can not be continued. 
Thus we regard as $0$ continuing endlessly after the first $0$ 
in the future and before the last $0$ in the past, 
if $0$ appears in the sequence.
For example, a triple collision orbit is represented as
\begin{equation}
{\bf n} = 
\dots 000n_{-l}n_{-(l-1)}\dots n_{-1} \cdot n_{0} n_{1} n_{2} \dots,
\end{equation}
or
\begin{equation}
{\bf n} = 
\dots n_{-3}n_{-2} n_{-1} \cdot n_{0}n_{1} \dots n_{m} 000 \dots,
\end{equation}
or
\begin{equation}
{\bf n} = 
\dots 000 n_{-l}n_{-(l-1)}\dots n_{-1} 
\cdot n_{0} n_{1} \dots n_{m} 000 \dots,
\end{equation}
where $n_{i}\in {\bf A}_{0}$. 
The shift operator $\sigma$ on ${\bf A}^{{\bf Z}}$ is defined by 
\begin{equation}
\sigma(\dots n_{-2}n_{-1} \cdot n_{0} n_{1} n_{2} \dots) =
\dots n_{-2}n_{-1} n_{0} \cdot n_{1} n_{2} \dots .
\end{equation}
We sometimes call finite symbol sequence the {\em word}. 
For instance, $12$ and $1211$ are words. 
\section{Triple collision manifold} \label{sec3}
The triple collisions are, in general, essential singularities 
and thus are not regularized. They are rare events. 
Therefore, it is hard to visualize the triple collisions. 
In order to investigate the structure near the triple collision, 
technical method is required. 
For celestial problem, McGehee has developed such method(McGehee, 1974). 
He has derived the equations of motion for the flow 
just on the triple collision. 
Its derivation is successive application of tricky transformations 
to the equations of motion Eq.(\ref{eq:H0}) 
and the energy conservation relation ${\cal H}=T({\bf p})-U({\bf q})=E$. 
A manifold on which the orbit experiences just the triple collision is 
obtained by setting the moment of inertia to be zero
(i.e., just triple collision) 
in the final energy conservation relation. 
This substitution is meaningful owing to tricky transformations
(i.e., scalings and time-transformations). 
This manifold is called the {\em triple collision manifold}(TCM). 
Thanks to similarity between celestial problem and Coulomb problem, 
this method can be also applied to our Coulomb problem. 
Since the transformations are tricky and complicated, 
in order to be self-contained, we show the derivation of it rapidly 
in the Appendix. 
For the detail of the derivation, 
the readers are recommended to consult with the article(McGehee, 1974). 
Starting from the energy conservation Eq.(\ref{eq:H0}) and 
the equations of motion for Eq.(\ref{eq:H0}),  
after lengthy calculation( i.e., six times of changes of variables ), 
we obtain the energy conservation relation with $r=0$, i.e., 
the moment of inertia $r^{2}=q_{1}^{2}+\xi q_{2}^{2} + q_{3}^{2}$ is zero
(just the triple collision) or with alternately the total energy $E=0$
\footnote{$r$ and $E$ appears as a term $rE$ in the energy relation 
and equations of motion(See the Appendix.). Therefore, 
the TCM and the flow on the TCM are the same for either $r=0$ and $E=0$.} : 
\begin{equation}
w^2+s^2 -1 + (1-s^2)^{2} W(s)^{-1} v^{2} = 0.
\label{eq:TCM}
\end{equation}
Equation (\ref{eq:TCM}) defines a surface in $(s,v,w)$-coordinates, 
i.e., the TCM.
The TCM is topologically equivalent 
to a sphere with four holes. 
Schematic picture for the TCM is depicted in Fig. \ref{fig1}.
For the definition of $s,v,w,W(s)$, see the Appendix. 
We only explain what $s$ represents. 
$s$ represents the configuration of three particles. 
$s$ is valued in $[-1,1]$. 
$s=-1$($s=1$) corresponds to the double collision 
between the particles $1$ and $2$($2$ and $3$), respectively.
The flow on the TCM is determined by the following equations of motion
\begin{eqnarray}
\frac{dv}{d\tau} & = & \frac{\lambda}{2} W(s)^{1/2} 
\left [ 1 - \frac{1-s^2}{W(s)} v^{2} \right ], \nonumber \\
\frac{ds}{d\tau} & = &  w,  \nonumber \\
\frac{dw}{d\tau} & = & -s + \frac{2s(1-s^2)}{W(s)}v^{2} 
+ \frac{1}{2} \frac{W'(s)}{W(s)}(1-s^2-w^2) 
- \frac{\lambda}{2}\frac{1-s^2}{W(s)^{1/2}}v w. 
\label{eq:flow_on_TCM}
\end{eqnarray}
For the definitions of $\lambda$,$s$,$v$,$w$, and $\tau$, 
see the Appendix. 
There are two critical points $c=(0,-v_{c},0)$ and $d=(0,v_{c},0)$, 
two infinite arms($a$ and $b$), 
and two infinite legs($e$ and $f$), 
where $v_{c}=W(0)^{\frac{1}{2}}$. 
Most of orbits on the TCM comes 
from infinity of one of two legs winding around it 
and goes out to infinity of one of two arms winding around it. 

There is an important property for the orbits with $r\neq 0$
(i.e., for total flow). 

{\bf Property 3.1}: 
For ${\cal H}=E<0$, the orbits runs inside of the TCM, 
i.e., $w^2+s^2 + (1-s^2)^{2} W(s)^{-1} v^{2} \leq 1$. 
On the other hand, for ${\cal H}=E>0$, 
the orbits runs outside of the TCM, 
i.e., $w^2+s^2 + (1-s^2)^{2} W(s)^{-1} v^{2} \geq 1$. 

Here after we consider the case of ${\cal H}=E<0$. 

If the orbit crosses $s=-1$($s=1$), the corresponding trajectory 
in the configuration space experiences the double collision 
between the particles $1$ and $2$($2$ and $3$) 
with the symbol $1$($2$), respectively.
After some collisions they are going out to infinity 
by changing its binary or not.
There also exist the triple collision orbits. 
By definition, 
the triple collision orbits are the orbits 
which start/end at the TCM(i.e., $r=0$). 
This means that the triple collision orbit can not be regularized, 
that is, the triple collision orbits can not continue after/before 
the triple collisions. 
As shown by McGehee(McGehee, 1974), 
the triple collision orbits form one parameter family. 
This fact is understood from the stability analysis 
of the critical point $c$ and $d$. 
The critical points $c$ and $d$ are the fixed points 
of the flow Eq.(\ref{eq:flow_on_TCM}). 
At the same time, they are the fixed points of 
the total flow Eq.(\ref{eq:flow}) which is not restricted to $r=0$. 
The stability analysis of the fixed points $c$ and $d$ 
shows that $\mbox{dim}({\cal W}^{s}(c))=2, \mbox{dim}({\cal W}^{u}(c))=1$ 
and $\mbox{dim}({\cal W}^{s}(d))=1, \mbox{dim}({\cal W}^{u}(d))=2$, 
where ${\cal W}^{s}(x)$ and ${\cal W}^{u}(x)$ are 
the stable and unstable manifolds of $x$, respectively. 
In Fig.\ref{fig2}, the schematic picture of 
${\cal W}^{s}(c), {\cal W}^{u}(c), {\cal W}^{s}(d)$, and ${\cal W}^{u}(d)$ 
is depicted. 
One branch of ${\cal W}^{s}(c)$(we call it ${\cal W}^{s,*}(c)$) 
comes into $c$ on the TCM 
from the outside of the TCM(i.e., $r \neq 0$) along the $v$-axis. 
Similarly, 
one branch of ${\cal W}^{u}(d)$(we call it ${\cal W}^{u,*}(d)$) 
goes out $d$ on the TCM 
to the outside of the TCM(i.e., $r\neq 0$) along $v$-axis. 
The other branch of ${\cal W}^{s}(c)$ and ${\cal W}^{u}(c)$
(we call them ${\cal W}_{TCM}(c)$) 
runs on the TCM and winds around the TCM. 
Similarly, 
the other branch of ${\cal W}^{u}(d)$ and ${\cal W}^{s}(d)$
(we call them ${\cal W}_{TCM}(d)$) 
runs on the TCM and winds around the TCM. 
In the outside of the TCM(i.e., $r \neq 0$) 
near the critical point $c$ and $d$, 
there exists one parameter family of the orbits 
which approach to ${\cal W}_{TCM}(c)$ as $\tau \rightarrow \infty$.
Similarly, there exists one parameter family of the orbits
which escape from ${\cal W}_{TCM}(d)$ as $\tau \rightarrow \infty$. 
Therefore, ${\cal W}_{TCM}(c)$ and ${\cal W}_{TCM}(d)$ determine 
the behaviour of the triple collision orbits. 
Unfortunately, this discussion here is limited to the neighbourhood of 
the critical points $c$ and $d$(i.e., local property). 
In the next section, we investigate 
${\cal W}_{TCM}(c)$ and ${\cal W}_{TCM}(d)$ numerically 
to show the global topological property of the triple collision orbits.  
For later use, we call the part of the TCM between two critical points 
$c$ and $d$ {\em the body of the TCM}, in precise, 
$\{ (s,v,w); -v_{c} \leq v \leq v_{c}, (s,v,w) \; \mbox{on the TCM} \}$. 

Our interest in this section is focused 
on the $\xi$-dependence of the flow on the TCM. 
In the celestial problem in collinear configuration 
considered in the section 10 of the article(McGehee, 1974), 
it is shown that for the case of $m_{1}=m_{3}=m$ and $m_{2}=\epsilon m$, 
when $\epsilon \rightarrow 0$, the orbits on the TCM wind around 
the TCM {\em infinitely}. 
Thanks to the similarity between celestial problem and Coulomb problem, 
for our Coulomb systems, the same argument is easily shown 
following the discussion of the article(McGehee, 1974). 

{\bf  Property 3.2:}When $\xi \rightarrow 0$, 
the orbits on the TCM wind around the body of the TCM {\em infinitely} 
often. 

This is due to the fact 
that $\frac{dv}{d\tau} \rightarrow 0$ as $\xi \rightarrow 0$. 
The proof is the same as the Proposition 10.1 of the article(McGehee, 1974).

In order to show the $\xi$-dependence of the flow on the TCM for large $\xi$,  
we numerically calculate ${\cal W}_{TCM}(c)$ and ${\cal W}_{TCM}(d)$. 
For this numerical integration of Eq.(\ref{eq:flow_on_TCM}), 
we used the fourth order Runge-Kutta method. 
Figure \ref{fig3} depicts ${\cal W}_{TCM}(c)$ and ${\cal W}_{TCM}(d)$ 
for $(Z,\xi) = (1,1)$. 
Figures \ref{fig4}(a) and (b) depict 
one branch(the unstable manifold) of ${\cal W}_{TCM}(c)$ 
for the cases $(Z,\xi)=(1,0.1)$ and $(1,6)$, respectively. 
The change of ${\cal W}_{TCM}(c)$ w.r.t. $\xi$ is clearly seen 
in Fig.\ref{fig4}. 
When $\xi$ is increased, 
the winding number ${\cal N}$ of ${\cal W}_{TCM}(c)$ or ${\cal W}_{TCM}(d)$
around the body of the TCM is decreased. 
Although further $\xi$ is increased from $\xi=6$, 
further change does not occur. 
Thus, when $\xi$ is increasing, 
${\cal N}$ is monotonically decreasing to a certain value. 
This property is also observed for different values of $Z$. 

{\bf Remark}: For our Coulomb system, the potential part of the Hamiltonian 
does not include the masses. Thus, $v_{c}$ is independent of $\xi$. 
For the gravitational case, 
the TCM has the same topology compared with our Coulomb system. 
But the potential part of the Hamiltonian includes the masses. 
This makes difference in the behaviour of 
the flow on the TCM. If $m_{1}=1,m_{2}=\xi$, and $m_{3}=1$, 
$v_{c} \sim \sqrt{\xi}$ as $\xi \rightarrow \infty$. 
Therefore, when $\xi \rightarrow \infty$, 
the winding number of ${\cal W}_{TCM}(c)$ or ${\cal W}_{TCM}(d)$ 
may not decrease as well as in the case of our Coulomb system. 
Thus, it is supposed that 
in the gravitational three-body problem in collinear configuration, 
the system hardly becomes hyperbolic, 
namely it always has stable orbits(i.e., tori). 

To summarize, when $\xi \rightarrow 0$, 
the orbit on the TCM wind around the body of the TCM 
{\em infinitely} often. 
When $\xi$ is increased, 
the number of times that the orbit winds around the body of TCM 
is monotonically decreasing to a certain value. 
The latter fact is related to what we examine in the next section. 
\section{Symbolic dynamics}\label{sec4}
In this section, 
we examine the global structure of the Poincar\'{e} surface of section 
by using the associate symbolic dynamics. 
The Poincar\'{e} surface of section is defined as follows: 
We denote the position on the line $Q_{1}=Q_{2}$ in the $(Q_{1},Q_{2})$-plane 
by $R$, i.e., $R=Q_{1}=Q_{2}$. On this line, the momenta $P_{1}$ and 
$P_{2}$ are specified by new variables $\theta$ and $R$.  
\begin{eqnarray}
P_{1} & = & \sqrt{T}\cos\theta + \sqrt{\frac{\xi T}{\xi+2}} \sin\theta ,
\nonumber \\
P_{2} & = & \sqrt{T}\cos\theta - \sqrt{\frac{\xi T}{\xi+2}} \sin\theta ,
\label{eq:PS}
\end{eqnarray}
where $T$ is the kinetic part of the total energy. 
Our surface of section is the $(\theta,R)$-plane 
where $0\leq R \leq R_{max}$, $0\leq \theta < 2\pi$ and 
$R_{max} = 2 - \frac{1}{2Z}$.
We denote this plane by ${\cal D}$. 
We define the map $\chi$ from the point ${\bf z}\in {\cal D}$ 
to the bi-infinite symbol sequence ${\bf n}$ 
which is an itinerary of the orbit 
started from ${\bf z}$ at the time zero. 
\begin{equation}
\begin{array}{cclll}
\chi & : & {\cal D} & \rightarrow & {\bf A}^{\bf Z} \\
     &   & {\bf z}  & \mapsto     &{\bf n}= \dots n_{-2}n_{-1} 
\cdot n_{0}n_{1}n_{2}\dots
\end{array}
\end{equation}
We define the map $\chi^{(+)}$ from the point ${\bf z}$ 
to the semi-infinite symbol sequence ${\bf n}^{(+)}$.
\begin{equation}
\begin{array}{cclll}
\chi^{(+)} & : & {\cal D} & \rightarrow & {\bf A}^{\bf N} \\
     &   & {\bf z}  & \mapsto     &
{\bf n}^{(+)}= n_{-1} \cdot n_{0}n_{1}n_{2}\dots
\end{array}
\end{equation}
We also consider the following map $\Xi_{l}$ 
from the semi-infinite symbol sequence 
to the finite symbol sequence with length $l+2$:
\begin{equation}
\Xi_{l}(n_{-1}\cdot n_{0} n_{1} \dots) = 
n_{-1} \cdot n_{0} n_{1} \dots n_{l}.
\end{equation}
Let us introduce 
\begin{eqnarray}
{\cal D}_{1} & = & 
\{ (\theta,R): 0 \leq\theta < \pi, 0\leq R \leq R_{max} \}, \nonumber \\
{\cal D}_{2} & = & 
\{ (\theta,R): \pi \leq\theta <2\pi, 0\leq R \leq R_{max} \}.
\end{eqnarray}
There are two symmetry operations on ${\cal D}$.
(1) if the orbit starting $ {\bf z}=(\theta,R) \in {\cal D}_{1}$ has 
the symbol sequence $ \dots m_{-2} m_{-1}\cdot n_{0} n_{1} \dots$, 
then $\dots n_{1}n_{0}\cdot m_{-1} m_{-2} \dots$ is a symbol sequence for 
the point $(2\pi - \theta, R) \in {\cal D}_{2}$. 
(2) if $\cdot n_{1} n_{2} n_{3} \dots$ is the future symbol sequence 
corresponding to the orbit starting at $(\theta,R)\in{\cal D}_{1}$ 
and if $\cdot m_{1} m_{2} m_{3} \dots$ is the future symbol sequence 
corresponding to the orbit starting at $(\pi-\theta,R)\in{\cal D}_{1}$,
then $\dots m'_{3} m'_{2} m'_{1} \cdot n_{1} n_{2} n_{3} \dots$ 
is the bi-infinite sequence corresponding to the full orbit starting 
at the point $(\theta,R) \in {\cal D}_{1}$, where $m'_{i}=1$ if $m_{i}=2$ 
and $m'_{i}=2$ if $m_{i}=1$. 
Thanks to these two symmetry operations, 
in order to investigate the global structure of ${\cal D}$, 
it is sufficient to study only the future orbits 
for the points in ${\cal D}_{1}$. 

We numerically construct the map $\chi^{(+)}$ in the following way: 
we consider the rectangle lattice whose lattice size is $1800 \times 1000$ 
for ${\cal D}_{1}$. 
For each lattice point ${\bf z}_{nm}, 1\leq n \leq 1800, 1\leq m \leq 1000$,  
we numerically obtained the truncated symbol sequence 
$\Xi_{60}(\chi^{(+)}({\bf z}_{nm}))$ 
by integrating the equations of motion for the Hamiltonian Eq.(\ref{eq:H}) 
using the sixth order symplectic integrator method 
as mentioned in \S \ref{sec2}. 

As established for celestial problem(Tanikawa and Mikkola, 2000),
the following properties are also true for our Coulomb system:

{\bf Property 4.1:} A trajectory in the $(Q_{1},Q_{2})$-plane 
transversely crosses the line $Q_{1}=Q_{2}$ except 
at $(\theta,R)=(0,0)$, if it does at all. 

{\bf Property 4.2:} If a trajectory crosses the line $Q_{1}=Q_{2}$ 
on the $(Q_{1},Q_{2})$-plane, a double collision occurs 
before the trajectory again crosses it.

The proof for the above Properties is 
the same as in the article(Tanikawa and Mikkola, 2000). 

From the latter Property 4.2, 
if we put the initial condition ${\bf z}$ in ${\cal D}_{1}$ 
whose trajectory crosses the line $Q_{1}=Q_{2}$,  
then the corresponding orbit has the symbol sequence 
$\dots n_{-3} n_{-2} 1 \cdot 2 n_{1}n_{2} \dots$, 
where $n_{i}\in {\bf A}$. 

The following two things are numerically checked for our Coulomb systems
(i.e., the case of $(Z,\xi)=(1,1)$ and the case of the helium)  
as observed for celestial problem(Tanikawa and Mikkola, 2000): 
The plane ${\cal D}_{1}$ is divided into two regions 
of the points ${\bf z}$ 
having $\Xi_{1}(\chi^{(+)}({\bf z}))=1\cdot 22$ and $1\cdot 21$.
Furthermore, it divided into four regions of the points 
${\bf z}$ having $\Xi_{2}(\chi^{(+)}({\bf z})) = 
1 \cdot 222$, $1 \cdot 221$, $1 \cdot 212$ and $1 \cdot 211$. 
This procedure is repeatedly applied. 
Then ${\cal D}_{1}$ is divided into smaller regions.  
The second thing is 
that the boundaries of regions of different symbol sequences form curves 
in ${\cal D}_{1}$. 
It is turned out that these curves are initial conditions of orbits 
which end in triple collision. 
So we call these curves the {\em triple collision curves}. 
We denote them by ${\cal C}_{TC}$. 

Figures \ref{fig5} and \ref{fig6} depict 
the triple collision curves ${\cal C}_{TC}$ 
for the case of $(Z,\xi)=(1,1)$ and the case of the helium, respectively. 
For the case of $(Z,\xi)=(1,1)$(Fig.\ref{fig5}), 
there is a torus in ${\cal D}_{1}$. 
The simple stable orbits(the torus)
may correspond to the Schubart orbits 
in celestial problem(Schubart, 1956). 
Therefore, we call these stable orbits the S-orbits.
In the context of atomic physics, 
it is usually called the asymmetric stretch orbit 
which was found by Simonovic and Rost(Simonovic and Rost, 2001). 
In Fig.\ref{fig5}, the region of the torus(S-orbits) 
is shown as a triangle area 
located at $\theta = \frac{\pi}{2}$. 
For the point ${\bf z}$ in this triangle area, 
it is that $\chi^{(+)}({\bf z}) = 1\cdot (21)^{\infty}$. 
In Fig.\ref{fig5}, 
we show the regions of the points ${\bf z}$ 
whose symbol sequence is 
$\Xi_{6}(\chi^{(+)}({\bf z}))=1\cdot 2n_{1}n_{2}\dots n_{6}$, 
where $n_{i} \in {\bf A}_{0}(i=1,2,\dots,6)$. 
There are the missing regions of the points ${\bf z}$ 
whose symbol sequence $\Xi_{l}(\chi^{(+)}({\bf z}))$ 
includes the words $1122$ and $2211$. 
Our numerical calculation up to the word length $15$ showed 
that the inadmissible words are only the words 
including the words $1122$ and $2211$. 
Note that $1122$ and $2211$ are also inadmissible words 
for the celestial problem(the case of $(Z,\xi)=(-1,1)$) 
and other inadmissible words exist
(Tanikawa and Mikkola, 2000). 
On the other hand, for the case of the helium(Fig.\ref{fig6}), 
i.e., large $\xi$, there is no torus. 
Figure \ref{fig6} shows the regions of the points ${\bf z}$ 
whose symbol sequence is 
$\Xi_{6}(\chi^{(+)}({\bf z}))=1\cdot 2n_{1}n_{2}\dots n_{6}$, 
where $n_{i} \in {\bf A}_{0}(i=1,2,\dots,6)$. 
In this case, there is no missing regions(i.e., no inadmissible words). 
When the symbol sequences $1.2n_{1}\dots n_{l}$ with length $l+2$ 
are considered, ${\cal D}_{1}$ is divided into $2^{l}$ partitions. 

In order to examine the hyperbolicity of the system, 
we investigate the foliated structure of ${\cal D}_{1}$.
Using the second symmetry of ${\cal D}_{1}$, 
we construct the triple collision curves 
whose orbit is started at triple collision in past. 
We denote the triple collision curves obtained in this way 
by ${\cal C}^{t}_{TC}$.
In Figs.\ref{fig7} and \ref{fig8}, 
we depict the triple collision curves 
whose orbit is started and/or ended at triple collision 
for the case of $(Z,\xi)=(1,1)$ and the case of the helium, respectively. 
Figs.\ref{fig7} and \ref{fig8} are constructed 
from Figs.\ref{fig5} and \ref{fig6}. 
It is clearly seen that for Fig.\ref{fig8} ${\cal C}_{TC}$ and 
${\cal C}^{t}_{TC}$ transversely cross each other, 
while for Fig.\ref{fig7} ${\cal C}_{TC}$ and ${\cal C}^{t}_{TC}$ 
transversely cross each other except near the torus. 
Since the dynamics of our system is continuous, 
we expect that 
${\cal C}_{TC}$ and ${\cal C}_{TC}^{t}$ 
do not cross the stable and unstable manifolds. 
For discontinuous system, such as billiard system, 
this is not the case. 
In fact, it is numerically confirmed that 
when $\xi$ is sufficiently large, 
the triple collision curves ${\cal C}_{TC}$(${\cal C}^{t}_{TC}$) is 
parallel to the stable(unstable) manifolds 
in the Poincar\'{e} section for the Poincar\'{e} map, respectively. 
Therefore, Fig.\ref{fig8}(sufficiently large $\xi$) 
manifests that the dynamics of the helium 
in collinear eZe configuration is hyperbolic. 
With some parameter values when the torus exists(for small value of $\xi$), 
${\cal C}_{TC}$ and ${\cal C}^{t}_{TC}$ do not foliate. 
In this case, it is observed that the tangency of 
the ${\cal C}_{TC}$ and ${\cal C}_{TC}^{t}$. 
This may manifest the tangency of the stable and unstable manifolds. 

The relationship between the observation in the previous section 
and the observation in Figures \ref{fig7} and \ref{fig8} 
is unclear at present. 
Next we clarify this relation 
by transforming the Poincar\'{e} plot in $(\theta,R)$-coordinates 
into that in $(s,v,w)$-coordinates. 
The Poincar\'{e} surface section in $(s,v,w)$-coordinates 
corresponding to that in $(\theta,R)$-coordinates  
is just the plane $s=0$. 
The transformation from $(\theta,R)$ to $(v,w)$ 
is as follows:
\begin{eqnarray}
v & = & \frac{1}{^{4}\sqrt{2}} 
\left ( P'_{1}+P'_{2} \right )\sqrt{2 - \frac{1}{2Z}-R} , 
\nonumber \\
w & = & ^{4}\sqrt{2} 
\left [2 \sqrt{2} \left ( 2 - \frac{1}{2Z} \right )   \right ]^{-\frac{1}{2}}
{\bf s}^{T}A^{T}{\bf p}' \sqrt{2 - \frac{1}{2Z}-R},
\end{eqnarray}
where
\begin{eqnarray}
P'_{1} & = & \cos \theta + \sqrt{\frac{\xi}{\xi+2}} \sin \theta, \nonumber \\
P'_{2} & = & \cos \theta - \sqrt{\frac{\xi}{\xi+2}} \sin \theta, 
\end{eqnarray}
and
\begin{equation}
{\bf p}' =  
\left (
\begin{array}{c}
-P'_{1} \\
P'_{1}-P'_{2} \\
P'_{2}
\end{array}
\right ), 
\hspace*{0.5cm}
{\bf s} = 
\frac{1}{\sqrt{2}}
\left (
\begin{array}{c}
-1 \\
0  \\
1
\end{array}
\right ). 
\end{equation}
Figure \ref{fig9} shows the Poincar\'{e} plot of 
the triple collision orbits in $(v,w)$-coordinates, 
which corresponds to Figure \ref{fig7} in $(\theta,R)$-coordinates 
for $(Z,\xi)=(1,1)$. 
$w\geq 0$($w \leq 0$) corresponds to ${\cal D}_{1}$(${\cal D}_{2}$), 
respectively. 
First, we note that 
there is a special solution along the line $s=w=0$, 
which starts from the critical point $d$ and ends in $c$. 
It is given by
\begin{eqnarray}
v & = & - v_{c} \mbox{tanh} 
\left ( \frac{\lambda}{2}(\tau - \tau_{0})\right ),
\nonumber \\
r & = & \frac{v_{c}^{2}}{2} \mbox{sech}^{2}
\left ( \frac{\lambda}{2}(\tau - \tau_{0})\right ),
\end{eqnarray}
where $\tau_{0}$ is determined by the initial condition.
This solution shows that ${\cal W}^{s,*}(c)$ and ${\cal W}^{u,*}(d)$ 
are degenerated. 
This type of solution also appears in symmetric gravitational 
four-body problem(Sekiguchi and Tanikawa, 2002). 
Second remarkable point is that the triple collision curves 
${\cal C}_{TC}$ and 
its reversed pair ${\cal C}_{TC}^{t}$ accumulate 
at ten points on the TCM(i.e., $w^{2} + W(0)^{-1}v^{2}=1$). 
Figure \ref{fig10} for $(Z,\xi)=(1,1)$ 
shows one branch of ${\cal W}_{TCM}(c)$.  
As shown in Fig.\ref{fig10}, 
these points are the points at which ${\cal W}_{TCM}(c)$ and 
${\cal W}_{TCM}(d)$ cross the plane $s=0$. 
We denote these points by ${\cal P}_{TCM,r=0}$.
It is easily understood that 
the number of points of ${\cal P}_{TCM,r=0}$ is related to 
the existence of the tori in the Poincar\'{e} surface of section 
$s=0$. 
If the tori exists, 
its outer most torus has periodic points. 
These periodic points have the stable and unstable manifolds. 
Some branches of these stable and unstable manifolds run 
toward ${\cal W}_{TCM}(c)$ and ${\cal W}_{TCM}(d)$. 
Therefore, the number of the points of ${\cal P}_{TCM,r=0}$ 
is directly related to the period of the periodic points 
associated to the tori. 
At the same time, the number of the points of ${\cal P}_{TCM,r=0}$ 
just corresponds to the winding number ${\cal N}$ of ${\cal W}_{TCM}(c)$ 
or ${\cal W}_{TCM}(d)$ around the body of the TCM 
as observed in the previous section. 
In Fig.\ref{fig11}, the case of $(Z,\xi)=(1,7)$ is shown. 
As the result of the previous section, 
when $\xi$ is large enough, 
the winding number of ${\cal W}_{TCM}(c)$ and ${\cal W}_{TCM}(d)$ 
saturates to certain value. 
In other words,
when $\xi$ is large enough, 
the number of the points of ${\cal P}_{TCM,r=0}$ also saturates 
to certain value which is, in fact, six(i.e., no torus case). 
Thus, the existence of torus in the Poincar\'{e} section $s=0$ 
is monitored by the number of points of ${\cal P}_{TCM,r=0}$.  
When $\xi$ is large enough, 
the triple collision curves ${\cal C}_{TC}$ 
transversely cross ${\cal C}_{TC}^{t}$. 
This would be a strong evidence of hyperbolicity of 
the system with large $\xi$, 
since it is numerically confirmed 
that ${\cal C}_{TC}$(${\cal C}_{TC}^{t}$) is parallel to 
the stable(unstable) manifold for the Poincar\'{e} map, respectively. 

The critical value of $\xi_c(Z)$ 
at which the winding number ${\cal N}$ is minimized, is calculateted. 
In Table \ref{table0}, we summarize the result. 
For $Z=1,2,3,4,5$, the critical value $\xi_{c}(Z)$ is order of $O(10)$. 
These critical value is numerically obtained. 
Thus, these are not true critical values.
However, we expect that at these true critical value 
the unstable manifold of $c$ 
and the stable manifold of $d$ degenerate. 
At these critical value $\xi_{c}(Z)$, the tori disappear.
At present, we do not know precise mechanism of 
this disapearance of the tori, e.g., even whether KAM-scenario 
is applied or not.

From the above numerical observation, we can state a conjecture. 
{\bf For the system Eq.(\ref{eq:H0}) with $(Z,\xi)$, 
there exists the critical value $\xi_{c}(Z)$ 
such that for $\xi > \xi_{c}(Z)$ the system is hyperbolic. }

Since for actual two-electron atom or ions 
the mass ratio is large, i.e., $m_{p}/m_{e}\approx 1840$, 
this means that the classical dynamics of 
H$^{-}$,He,Li$^{+}$,Be$^{2+}$,$\dots$ 
in collinear eZe configuration is hyperbolic. 
For inifinite mass ratio, 
it was numerically shown that the symbolic description is complete and 
the dynamics probably hyperbolic(Bai, Gu and Yuan, 1998). 

\subsection{The case of $(Z,\xi)=(1,1)$} \label{sec4-1}
In this subsection, 
we investigate the detailed structure of the Poincar\'{e} surface of 
section for the case of $(Z,\xi)=(1,1)$. 
In this case, there is a torus whose symbol sequence 
is $1\cdot 21212\dots$ as shown in Figs.\ref{fig5} and \ref{fig7}. 
This torus has the periodic points with period $6$ 
as the outermost part. 
Note that we include the periodic points in ${\cal D}_{2}$ 
and count the period. 
These periodic points $\alpha,\beta,\gamma$ has 
stable and unstable manifolds. 
$\alpha:(\theta,R)=(0.5\pi,1.385)$, $\beta:(\theta,R)=(0.5292\pi,1.275)$,and 
$\gamma:(\theta,R)=(0.4708\pi,1.275)$.
From the numerical calculation, 
one branch of ${\cal W}^{s}(\alpha)$ is equal to one branch of 
${\cal W}^{u}(\beta)$. 
In the same way, 
one branch of ${\cal W}^{s}(\beta)$ is equal to one branch of 
${\cal W}^{u}(\gamma)$ 
and one branch of ${\cal W}^{s}(\gamma)$ is equal 
to one branch of ${\cal W}^{u}(\alpha)$. 
We can construct the stable manifolds of $\alpha,\beta$, and $\gamma$ 
outside the S-orbits by examining the long orbits with the symbol sequence 
$1\cdot 21212\cdots$. 
The result is depicted in Fig.\ref{fig12}(a). 
In Fig.\ref{fig12}(b), we also depict the unstable manifolds 
by using the second symmetry of ${\cal D}_{1}$. 
The stable manifolds of $\alpha$, $\beta$, and $\gamma$ 
outside S-orbits are basic boundaries of the partitions of ${\cal D}_{1}$. 

We further examined the symbol sequence 
$\Xi_{60}(\chi^{(+)}({\bf z}_{mn}))$ for each ${\bf z}_{mn}$ 
($0\leq m\leq 1800$,$0\leq n \leq 1000$). 
As a result, ${\cal D}_{1}$ is divided into, at least, ten partitions: 
$S$,$I_{i}(i=1,2,\dots,9)$. 
$S$ stands for the S-orbits. 
Other partitions $I_{i}(i=1,2,\dots,9)$ are defined as follows.
We examined the symbol sequences $\Xi_{60}(\chi^{(+)}({\bf z}))$ 
along the line $\theta = 0.2\pi$, $0.5\pi$, $0.55\pi$ and $0.7\pi$. 
For the line $\theta = 0.2 \pi$, 
we find that the symbol sequences are distributed 
as it decreases from $1\cdot (2)^{\infty}$ 
to $1\cdot (21)^{\infty}$ with increasing $R$. 
We divide symbol sequences into three groups: 
$1\cdot (2)^{\infty}$, $1\cdot(2)^{n}12\dots$, and $1\cdot(21)^{n}\dots$
($n\geq 2$). 
We call the regions with these symbol sequences $I_{7}$, $I_{4}$ and $I_{1}$.

For the line $\theta = 0.5\pi$, 
the symbol sequences are distributed 
as it increases from $1\cdot 2(1)^{\infty}$ to $1\cdot (21)^{\infty}$. 
We divide symbol sequences into three groups:
$1\cdot 2(1)^{\infty}$, $1\cdot 2(1)^{n}21\dots$, and 
$1\cdot(21)^{n}\dots$($n\geq 2)$. 
We call the regions with these symbols sequences $I_{8}$, $I_{5}$ and $I_{2}$. 

By examining the symbol sequences 
along the line $\theta = 0.55 \pi$ and $0.7\pi$, 
we find that along the line $R=1.3$ from $\theta = \pi$ to $0.5\pi$, 
the symbol sequences are distributed 
as it decreases from $1\cdot 21(2)^{\infty}$ to $1\cdot (21)^{\infty}$ 
with decreasing $\theta$. 
We divide symbol sequences into three groups:
$1\cdot 21(2)^{\infty}$, $1\cdot 21(2)^{n}12\dots$ and 
$1\cdot (21)^{n}\dots$($n\geq 2$). 
We call the regions with these symbols sequences $I_{9}$, $I_{6}$ and $I_{3}$.
Fig.\ref{fig13}(a) shows these partitions of ${\cal D}_{1}$. 
In Fig.\ref{fig13}(b), 
the time-reversed partitions by the second symmetry of ${\cal D}_{1}$ 
are shown with them of Fig.\ref{fig13}(a). 
In Table \ref{table1}, 
the characteristic feature of $S$ and $I_{i}(i=1,2,\dots,9)$ 
is summarized. 

Next we prepared many points in each partition and 
checked which partitions they are mapped to. 
In Table \ref{table2}, 
we summarized the transition among the partitions. 
From Table \ref{table2}, we drew the diagram of the 
transitions among partitions in Fig.\ref{fig14}. 
From this diagram, 
we know that typical orbits travel 
around the regions near the torus($I_{1}$,$I_{2}$, and $I_{3}$) 
and/or around the regions in which the orbits feel large instability
($I_{4}$,$I_{5}$ and $I_{6}$), and, that in most cases, 
they escape to the regions($I_{7}$, $I_{8}$ and $I_{9}$) 
except the S-orbits and non-wandering orbits(probably repellor). 
\subsection{The case of helium} \label{sec4-2}
For the case of helium, 
Figs. \ref{fig6} and \ref{fig8} tell all features of the partitions. 
The partitions formed by ${\cal C}_{TC}$ 
is, at first sight, expected to be the Markov partitions. 
If we obtain ${\cal C}_{TC}$ using the symbol sequences 
$1.2n_{1}\dots n_{l}$, then ${\cal D}_{1}$ is divided 
into $2^{l}$ partitions. 
From numerical observation, there is no inadmissible words. 
However, it is clear that $\bar{1}$ and $\bar{2}$ are removed, 
since these orbits correspond to fixed points at inifinity. 
The Markov partition should be shaped as parallelepiped. 
Thus probably, in mathematical rigorous sense, 
the partitions in our Poincar\'{e} section are not the Markov partitions. 
If we construct the Markov partitions in our Poincar\'{e} section, 
they would be infinite. 
This reflects the fact that the dynamics of helium atom 
in collinear eZe configuration exhibits intermittency
(Richter, Tanner and Wintgen, 1993, Tanner and Wintgen, 1995, 
Tanner, Richter and Rost, 2000). 
The intermittency observed in helium atom in collinear eZe configuration 
is the behaviour that the electrons are going back and forth 
between long flight(i.e., almost ionization) and short flight
(i.e., successive collisions with the nucleus). 
This intermittency is due to the existence of the fixed points at infinity. 
Its behaviour is very similar to hyperbola billiard(Sieber and Steiner, 1990). 
The transverse crossing of ${\cal C}_{TC}$ and ${\cal C}^{t}_{TC}$ 
strongly suggests that the dynamics is hyperbolic. 
\subsection{The Poincar\'{e} plots for other two-electron ions}
For the case of the inifinite nucleus mass, 
it was investigated by Bai et al(Bai et al, 1998). 
It was numerically shown that for $Z\geq 1$, the symbolic description is 
complete and it suggests that the system is hyperbolic. 
We now consider the finite mass case. 
The investigation above does not consider the $Z$-dependence 
of the dynamics of two-electron atom and ions in collinear eZe configuration. 
Now in order to strengthen our claim that the dynamics of them is 
fully chaotic, probably hyperbolic, 
we calculated the Poincar\'{e} plots for thirty atom and ions 
among the systems H$^{-}$, He, Li$^{+}$, ... , Fm$^{98+}$ 
in collinear eZe configuration. 
We do not plot them since there is no space. 
This calculation showed that even though $Z$ becomes large 
at the order $O(100)$ and then the interaction terms becomes small 
in Eq.(\ref{eq:H}), their large mass ratio $\xi$ overcomes the $Z$-dependence 
to result in the fact 
that the Poincar\'{e} plot is filled by chaotic sea except the escape region 
and there is no visible torus. 
These observations strengthen our claim. 
\section{Experimental aspect}\label{sec5}
From our observations, 
we know that 
the system with small mass ratio qualitatively differs 
from the system with large mass ratio, 
namely the existence of stable orbits. 
This difference would be experimentally observed. 
A possible candidates are the antiproton-proton-antiproton
($\overline{\mbox{p}}$-p-$\overline{\mbox{p}}$) system, 
the positronium negative ion(Pr-($e$-$\overline{e}$-$e$)).
which corresponds to the case of $Z=1,\xi=1$.
We neglect relativistic effects, bremsstrahlung and hyperfine interaction.
If this system has the bound states, 
the eigenenergies possesses the effects of the torus. 
Most convenient analysis is the Fourier transform of 
the density of states for the spectrum 
with the angular momentum ${\bf L}=0$,  
which gives the information on the length and the stability 
of periodic orbits. 
For the positronium negative ion, 
the EBK quantization was done(Simonovic and Rost, 2001).
Stable antisymmetric orbit was obtained and was quantized 
to explain the energy spectrum. 
From our observation, 
as a ZeZ configuration,  
the positive hydrogen molcule $H_{2}^{+}$ 
have stable periodic orbits. 

In addition, 
although large $Z$ two-electron ions are experimentally unrealistic, 
small $Z$ two-electron ions serve us the data 
for the manifestation of our finding
(i.e., hyperbolicity(strong chaotic property)). 
H$^{-}$ and He have been already analyzed by the semiclassical method
(Gaspard and Rice, 1993, Ezra et al., 1991). 
Therefore, Li$^{+}$ is another candidate. 
If we change $Z$, the degree of the intermitency 
of the dynamics would change(Tanner and Wintgen, 1995). 
This change may be reflected in the behaviour of 
the quantum defect of H$^{-}$, He, Li$^{+}$, Be$^{2+}$, .... 

\section{Summary}\label{sec6}

In this paper, 
we have investigated the bifurcation in classical Coulomb three-body problem 
in collinear eZe configuration with finite mass of all three particles. 
In particular, the main result is that 
when the mass ratio $\xi$ is changed, 
the change of the flow on the TCM is directly related to 
the existence(or collapse) of the tori. 
This result suggests that 
if the mass ratio $\xi$ is sufficiently large, 
the dynamics of these system is hyperbolic. 
This result is consistent with the result(i.e., infinite nucleus mass) 
by Bai et al(Bai, Gu and Yuan, 1998).
But by our analysis, 
it was shown that there surely exists a threshold value of $\xi$ 
for hyperbolicity. 
These threshold value is order of $O(10)$. 
Therefore, H$^{-}$, He, Li$^{+}$, Be$^{2+}$, ... 
in the collinear eZe configuration is hyperbolic. 
Since our analysis is based on numerical one, 
we do not have the proof of this fact. 
We call for rigorous proof of this result. 

\ack
The author is grateful to Prof. K.Tanikawa for stimulating discussion 
on three-body problem in celestial mechanics. 
\section*{Appendix}
In this Appendix, we derive the TCM for our Coulomb systems. 
Its derivation is completely parallel 
to the case of celestial problem. 
The readers are recommended to consult with the article for detail
(McGehee, 1974).
We start from the Hamiltonian Eq.(\ref{eq:H0}). 
\begin{equation}
{\cal H} = T({\bf p}) - U({\bf q}) = E.
\end{equation}
where we removed the dash in Eq.(\ref{eq:H0}).  
$T({\bf p})$ is the kinetic part of ${\cal H}$ and 
$-U({\bf q})$ is the potential part of ${\cal H}$.  
The equations of motion in the Cartesian coordinates 
are given by
\begin{eqnarray}
\frac{d{\bf q}}{dt} & = &  M^{-1}{\bf p}, \nonumber \\
\frac{d{\bf p}}{dt} & = &  \nabla U({\bf q}),
\end{eqnarray}
with the mass matrix
\begin{equation}
M = \left [
\begin{array}{ccc}
  1 &  0  & 0\\
  0 &  \xi & 0\\
  0 &  0  & 1
\end{array}
\right ].
\end{equation}
We set the center of mass to be zero.
\begin{equation}
q_{1} + \xi q_{2}+ q_{3}= 0.
\label{eq:center_of_mass}
\end{equation}
We set the total momentum to be also zero.
\begin{equation}
p_{1}+p_{2}+p_{3} = 0.
\label{eq:total_momentum}
\end{equation}
We consider the transformation from the Cartesian coordinates
$(q_{1},q_{2},q_{3},p_{1},p_{2},p_{3})$ with 
Eqs.(\ref{eq:center_of_mass}) and (\ref{eq:total_momentum}) 
to the McGehee's variables $(r,s,v,w)$. 
First, we set 
\begin{eqnarray}
r & = & ({\bf q}^{T}M{\bf q})^{1/2}, \nonumber \\
{\bf s} & = & r^{-1}{\bf q}.
\label{eq:app1}
\end{eqnarray}
$r^{2}$ is the moment of inertia. 
${\bf s}$ represents the configuration of three particles. 
The triple collision corresponds to $r=0$. 
Now we parameterize the variable ${\bf s}$ by single variable $s$. 
To do so, 
we set 
\begin{eqnarray}
{\bf a} & = & (a_1,a_2,a_3),\; a_1=a_2<a_3 \nonumber \\
{\bf b} & = & (b_1,b_2,b_3),\; b_1<b_2=b_3 
\end{eqnarray}
where
\begin{equation}
{\bf a}^{T}M{\bf a} = {\bf b}^{T}M{\bf b} = 1.
\end{equation}
Since we fix the center of mass, 
namely $a_{1}+\xi a_{2}+a_{3}=0$ and $b_{1}+\xi b_{2}+b_{3}=0$, 
then we have
\begin{eqnarray}
{\bf a} & = & 
\left ( 
- \frac{1}{\sqrt{(1+\xi)(2+\xi)}}, 
- \frac{1}{\sqrt{(1+\xi)(2+\xi)}},
\frac{1+\xi}{\sqrt{(1+\xi)(2+\xi)}}
\right ), \nonumber \\ 
{\bf b} & = & 
\left ( 
-\frac{1+\xi}{\sqrt{(1+\xi)(2+\xi)}}, 
\frac{1}{\sqrt{(1+\xi)(2+\xi)}},
\frac{1}{\sqrt{(1+\xi)(2+\xi)}}
\right ).
\end{eqnarray}
It is shown that the variable ${\bf s}$ 
is parametrized by the variable $s \in [-1,1]$ as follows
(See the article for detail(McGehee, 1974).): 
\begin{equation}
{\bf s} = (\sin(2\lambda))^{-1} \left [ \sin(\lambda(1-s){\bf a} + 
\sin(\lambda(1+s)) {\bf b} \right ],
\label{eq:s}
\end{equation}
with
\begin{equation}
\cos(2\lambda) = {\bf a}^{T}M{\bf b} = \frac{1}{1+\xi}.
\end{equation}
Here $\lambda$ is valued as $0\leq \lambda \leq \frac{\pi}{2}$. 
We denote the map from $s \in [-1,1]$ 
to the configuration ${\bf s}$ by $S$:
\begin{equation}
\begin{array}{cccl}
S: & [-1,1] & \rightarrow & 
\{{\bf s}: {\bf s}^{T}M{\bf s}=1, s_{1}+\xi s_{2}+ s_{3} = 0 \},\\
   & s & \mapsto & {\bf s} .
\end{array}
\end{equation}
Note that $s=-1$($s=1$) corresponds 
to the collision of the particles $1$ and $2$($2$ and $3$), respectively. 
$U({\bf s})$ becomes 
\begin{eqnarray}
V(s) & = & U(S(s))\nonumber \\
& = & \sin(2\lambda) \left [
\frac{1}{(b_2-b_1)\sin(\lambda(1+s))} + 
\frac{1}{(a_3-a_2)\sin(\lambda(1-s))} \right . \nonumber \\
& & \left . -
\frac{1}{Z\{(b_2-b_1)\sin(\lambda(1+s)) + (a_3-a_2)\sin(\lambda(1-s)) \}}
\right ]
\end{eqnarray}
Here we set two matrices $A_1$ and $A_2$ 
\begin{equation}
A_{1} = 
\left [
\begin{array}{rrr}
1 & 1 & 1 \\
1 & 1 & 1 \\
1 & 1 & 1 
\end{array}
\right ], \; 
A_{2} =
\left [
\begin{array}{rrr}
 0 &  1 & -1 \\
-1 &  0 &  1 \\
 1 & -1 &  0 
\end{array}
\right ], 
\end{equation}
and define the matrix $A$ 
\begin{equation}
A = \frac{1}{2+\xi} A_{1}M + 
\left ( \frac{\xi}{2+\xi} \right )^{\frac{1}{2}}
M^{-1} A_{2}.
\end{equation}
Now the variable $s$ is defined as the inverse of 
Eq.(\ref{eq:s}).
\begin{equation}
s = S^{-1}({\bf s}).
\label{eq:app2}
\end{equation}
We further set 
\begin{eqnarray}
v & = & r^{\frac{1}{2}}{\bf p}^{T}{\bf s}, \nonumber \\
w & = & r^{\frac{1}{2}}(1-s^{2}) W(s)^{-\frac{1}{2}} 
{\bf s}^{T}A^{T}{\bf p},
\label{eq:app3}
\end{eqnarray}
where
\begin{equation}
W(s) = 2 (1-s^{2})V(s).
\end{equation}
We also employ time-transformation two times.
\begin{equation}
dt = r^{\frac{3}{2}} dt'\; \mbox{and} \; 
dt' = \lambda (1-s^{2})W(s)^{-\frac{1}{2}} d\tau.
\label{eq:app4}
\end{equation}
In short, 
the change of variables 
Eqs.(\ref{eq:app1}),(\ref{eq:app2}),(\ref{eq:app3}), and (\ref{eq:app4}) 
give the wanted relations, 
i.e., the energy conservation and the equations of motion.
The energy conservation becomes 
\begin{equation}
1- \frac{2w^2}{1-s^2} = \frac{2(1-s^2)}{W(s)}(v^{2}-2rE) - 1.
\label{eq:manifold}
\end{equation}
The equations of motion become
\begin{eqnarray}
\frac{dr}{d\tau} & = & \frac{\lambda(1-s^2)}{W(s)^{1/2}} rv, \nonumber \\
\frac{dv}{d\tau} & = & \frac{\lambda}{2} W(s)^{1/2}
\left [
1- \frac{1-s^2}{W(s)}(v^2-4rE)
\right ], \nonumber \\
\frac{ds}{d\tau} & = & w, \nonumber\\
\frac{dw}{d\tau} & = & -s + \frac{2s(1-s^2)}{W(s)}(v^{2}-4rE) + 
\frac{1}{2} \frac{W'(s)}{W(s)}(1-s^2-w^2) \nonumber \\
& & - \frac{\lambda(1-s^2)}{2W(s)^{1/2}}vw. 
\label{eq:flow}
\end{eqnarray}
Finally, we set $r=0$. 
Thanks to the above transformations, this substitution is meaningful, 
since the singularities from the double collisions have been removed. 
From Eqs.(\ref{eq:manifold}) and (\ref{eq:flow}), 
we obtain Eqs.(\ref{eq:TCM}) and (\ref{eq:flow_on_TCM}). 
We can also set $E=0$. Then the same energy relation and 
the equations of motion are obatained, since $r$ and $E$ appear 
as a term $rE$. The dynamis of scattering flow for $E=0$ was 
investigated by Bai et al(Bai,Gu,Yuan, 1998) 
for the case of the inifinite nucleus mass. 
%
\section*{References}

\begin{harvard}


\item[]
{Bai Z-Q, Gu Y and Yuan J-M}  1998
{\it Physica D} {\bf 118} 17

\item[]
{Einstein A} 1917
{\it Verh.Dtsch.Phys.Ges.} {\bf 19} 82

\item[]
{Ezra G S, Richter K, Tanner G and Wintgen D} 1991
{\it J.Phys.B} {\bf 24} L413

\item[]
{Gaspard P and Rice S A} 1993
{\it Phys.Rev.A} {\bf 48} 54

\item[]
{Gutzwiller M C} 1971
{\it J.Math.Phys.} {\bf 12} 343

\item[]
{Gutzwiller M C} 1990
{\it Chaos in classical and quantum mechanics} 
(New York: Springer, New York)


\item[]
{McGehee R} 1974
{\it Invent. Math.} {\bf 27} 191

\item[]
{Mikkola S and Tanikawa K} 1999
{\it Celst.Mech.Dyn.Astron.} {\bf 74} 287

\item[]
{Mikkola S and Tanikawa K} 1999
{\it Mon.Not.R.Astron.Soc.} {\bf 310} 745

\item[]
{Siegel C L and Moser J K} 1971 
{\it Lectures on Celestial Mechanics} 
(Berlin: Springer-Verlag)

\item[]
{Poincar\'{e} H} 1899
{\it Les m\'{e}thodes nouvelles de la m\'{e}chanique c\'{e}leste I, II et III} 
(Paris: Gauthier-Villars)

\item[]
{Richter K, Tanner G and Wintgen D} 1993
{\it Phys.Rev.A} {\bf 48} 4182

\item[]
{Schubart J} 1956
{\it Astron.Nachr.} {\bf 283} 17

\item[]
{Sekiguchi M and Tanikawa K} 2002
{``On the Symmetric Collinear Four-Body Problem''} 
{\it preprint}

\item[]
{Sieber M and Steiner F} 1990
{\it Physica D} {\bf 44} 248

\item[]
{Simonovic N and Rost J-M} 2001
{\it Euro.Phys.J. D} {\bf 15} 155


\item[]
{Tanikawa K and Mikkola S} 2000
{\it Chaos} {\bf 10} 649

\item[]
{Tanner G, Richter K and Rost J-M} 2000
{\it Rev.Mod.Phys.} {\bf 72} 497

\item[]
{Tanner G and Wintgen D} 1995
{\it Phys.Rev.Lett.} {\bf 75} 2928



\item[]
{Yoshida H} 1990
{\it Phys.Lett.A} {\bf 150} 262







\end{harvard}

%
%
\newpage


\begin{figure}
\begin{center}
\leavevmode
\epsfxsize=10cm
\end{center}
\caption{The triple collision manifold: 
There are two critical points $c$ and $d$.}
\label{fig1}
\end{figure}

\begin{figure}
\begin{center}
\leavevmode
\epsfxsize=7cm
\end{center}
\caption{The schematic picture of the stable and unstable manifolds of 
the critical points $c$ and $d$: 
${\cal W}(c)$ and ${\cal W}(d)$ are depicted. 
$\mbox{dim}({\cal W}^{s}(c))=2$, 
$\mbox{dim}({\cal W}^{u}(c))=1$, 
$\mbox{dim}({\cal W}^{s}(d))=1$, 
$\mbox{dim}({\cal W}^{u}(d))=2$.
}
\label{fig2}
\end{figure}

\begin{figure}
\begin{center}
\leavevmode
\epsfxsize=12cm
\end{center}
\caption{The stable and unstable manifolds of the critical points $c$ 
and $d$ on the TCM for $(Z,\xi)=(1,1)$: 
${\cal W}_{TCM}(c)$ and ${\cal W}_{TCM}(d)$. 
Two circles indicate the positions of two critical points $c$ and $d$. }
\label{fig3}
\end{figure}

\begin{figure}
\begin{center}
\begin{tabular}{ll}
(a) & (b)\\
\leavevmode
\epsfxsize=7cm
&
\leavevmode
\epsfxsize=7cm
\end{tabular}
\end{center}
\caption{The unstable manifold of the critical point $c$ 
on the TCM: 
(a)$(Z,\xi)=(1,0.1)$, (b)$(Z,\xi)=(1,6)$.
}
\label{fig4}
\end{figure}

\begin{figure}
\begin{center}
\leavevmode
\epsfxsize=12cm
\end{center}
\caption{The triple collision curves ${\cal C}_{TC}$ 
for the case of $(Z,\xi)=(1,1)$: We also plot the torus region(triangle area). 
The curves are obtained by the symbol sequences with length $7$ in future.}
\label{fig5}
\end{figure}

\begin{figure}
\begin{center}
\leavevmode
\epsfxsize=12cm
\end{center}
\caption{The triple collision curves ${\cal C}_{TC}$ for the case of helium: 
The curves are obtained by the symbol sequences with length $7$ in future.}
\label{fig6}
\end{figure}

\begin{figure}
\begin{center}
\leavevmode
\epsfxsize=12cm
\end{center}
\caption{The triple collision curves ${\cal C}_{TC}$ and ${\cal C}^{t}_{TC}$
for the case of $(Z,\xi)=(1,1)$: We also plot the torus region(triangle area). 
These curves are obtained from the data of Fig.5. }
\label{fig7}
\end{figure}

\begin{figure}
\begin{center}
\leavevmode
\epsfxsize=12cm
\end{center}
\caption{The triple collision curves ${\cal C}_{TC}$ and ${\cal C}^{t}_{TC}$ 
for the case of the helium: These curves are obtained 
from the data of Fig.6. }
\label{fig8}
\end{figure}

\begin{figure}
\begin{center}
\leavevmode
\epsfxsize=12cm
\end{center}
\caption{The Poincar\'{e} plot of the triple collision orbits 
${\cal C}_{TC}$ and ${\cal C}_{TC}^{t}$ 
in $(v,w)$-coordinates for $(Z,\xi)=(1,1)$. 
It corresponds to Fig.7. }
\label{fig9}
\end{figure}

\begin{figure}
\begin{center}
\leavevmode
\epsfxsize=12cm
\end{center}
\caption{The Poincar\'{e} plot of the triple collision orbits 
and one branch of the unstable manifold of the critical point $c$ 
for $(Z,\xi)=(1,1)$. 
Three crossing points(including the critical point $c$) 
where the unstable manifold crosses the Poincar\'{e} section $s=0$ 
are indicated by squares. 
}
\label{fig10}
\end{figure}

\begin{figure}
\begin{center}
\leavevmode
\epsfxsize=12cm
\end{center}
\caption{The Poincar\'{e} plot of the triple collision orbits 
${\cal C}_{TC}$ and ${\cal C}_{TC}^{t}$ for $(Z,\xi)=(1,7)$ 
in $(v,w)$-coordinates. }
\label{fig11}
\end{figure}

\begin{figure}
\begin{center}
\begin{tabular}{ll}
(a) & (b) \\
\leavevmode
\epsfxsize=6cm
\leavevmode
\epsfxsize=6cm
\end{tabular}
\end{center}
\caption{The stable and unstable manifolds of $\alpha,\beta$, and $\gamma$ 
for the case of $(Z,\xi)=(1,1)$:
(a) The stable manifolds of $\alpha,\beta$, and $\gamma$, 
(b) The unstable manifolds(dotted lines) of $\alpha,\beta$ and $\gamma$ 
are also added to (a).}
\label{fig12}
\end{figure}

\begin{figure}
\begin{center}
\begin{tabular}{l}
(a) \\
\leavevmode
\epsfxsize=12cm
(b) \\
\leavevmode
\epsfxsize=12cm
\end{tabular}
\end{center}
\caption{The partitions in ${\cal D}_{1}$ for the case of $(Z,\xi)=(1,1)$: 
(a) The partitions in ${\cal D}_{1}$,
(b) The time-reversed partitions are also added to (a). }
\label{fig13}
\end{figure}

\begin{figure}
\begin{center}
\leavevmode
\epsfxsize=12cm
\end{center}
\caption{The transition diagram 
for the case of $(Z,\xi)=(1,1)$ 
among the partitions derived from Table 2.}
\label{fig14}
\end{figure}



\clearpage


\begin{table}
\begin{center}
\begin{tabular}{|l|l|}\hline
Z & $\xi_{c}(Z)$ \\ \hline \hline
1 & 6.4 \\ \hline
2 & 15.0 \\ \hline
3 & 23.5\\ \hline
4 & 32.0 \\ \hline
5 & 40.4 \\ \hline
\end{tabular}
\end{center}
\caption{The critical value $\xi_{c}(Z)$.}
\label{table0}
\end{table}

\begin{table}
\begin{center}
\begin{tabular}{|l|l|l|}\hline
Partition & Character & Symbol sequence \\ \hline \hline
$S$     & S-orbits(torus)&  $1\cdot (21)^{\infty}$\\ \hline
$I_{1}$ & & $1\cdot(21)^{n}\dots(n\geq 2)$\\ \hline
$I_{2}$ & & $1\cdot(21)^{n}\dots(n\geq 2)$\\ \hline
$I_{3}$ & & $1\cdot(21)^{n}\dots(n\geq 2)$\\ \hline
$I_{4}$ & & $1\cdot(2)^{n}12\dots(n\geq 2)$\\ \hline
$I_{5}$ & & $1\cdot 2(1)^{n}21\dots(n\geq 2)$\\ \hline
$I_{6}$ & & $1\cdot 21(2)^{n}12\dots(n\geq 2)$ \\ \hline
$I_{7}$ & Escape & $1\cdot (2)^{\infty}$ \\ \hline
$I_{8}$ & Escape & $1\cdot 2(1)^{\infty}$ \\ \hline
$I_{9}$ & It is mapped to $I_{7}$. & $1\cdot 21(2)^{\infty}$ \\ \hline
\end{tabular}
\end{center}
\caption{The characteristic feature of each partition for $(Z,\xi)=(1,1)$.}
\label{table1}
\end{table}

\begin{table}
\begin{center}
\begin{tabular}{l}\hline
Transitions  \\ \hline \hline
$S$      $\rightarrow$  $S$ \\ \hline
$I_{1}$  $\rightarrow$  $I_{2},I_{5},I_{8}$\\ \hline
$I_{2}$  $\rightarrow$  $I_{3},I_{6},I_{9}$\\ \hline
$I_{3}$  $\rightarrow$  $I_{1}$\\ \hline
$I_{4}$  $\rightarrow$  $I_{1},I_{2},I_{4},I_{5},I_{7},I_{8}$\\ \hline
$I_{5}$  $\rightarrow$  $I_{2},I_{3},I_{5},I_{6},I_{8},I_{9}$\\ \hline
$I_{6}$  $\rightarrow$  $I_{4}$\\ \hline
$I_{7}$  $\rightarrow$  Escape\\ \hline
$I_{8}$  $\rightarrow$  Escape\\ \hline
$I_{9}$  $\rightarrow$  $I_{7}$\\ \hline
\end{tabular}
\end{center}
\caption{The transitions among the partitions for $(Z,\xi )=(1,1)$.}
\label{table2}
\end{table}

\end{document}